\documentclass[a4paper,fleqn]{cas-dc}

\usepackage[numbers]{natbib}

\usepackage{tikz}
\usepackage{pgfplots}
\usepackage{pdfx}

\def\tsc#1{\csdef{#1}{\textsc{\lowercase{#1}}\xspace}}
\tsc{WGM}
\tsc{QE}

\begin{document}
\let\WriteBookmarks\relax
\def\floatpagepagefraction{1}
\def\textpagefraction{.001}

\shorttitle{Wavelet-Space Super-Resolution}    

\shortauthors{}  

\title [mode = title]{Wavelet-Space Representations for Neural Super-Resolution in Rendering Pipelines}  

\tnotemark[1] 

\tnotetext[1]{} 

%

\author[1]{Prateek Poudel}[
    orcid    = 0009-0006-1947-825X,
    linkedin = prateekpoudel
]
\cormark[1]
\fnmark[1]
\ead{prateekpoudel61@gmail.com}
\credit{Conceptualization, Methodology, Software (Unreal plugin), Writing – original draft, 
        Writing – review \& editing, Project structuring}

\author[1]{Prashant Aryal}[
    orcid    = 0009-0004-6756-6807,
    linkedin = pr-a-sh-ant
]
\fnmark[1]
\ead{aryalprashant07@gmail.com}
\credit{Data curation, Dataset extraction, Training, 
        Writing – original draft, Visualization}

\author[1]{Kirtan Kunwar}[
    orcid    = 0009-0001-7913-1497,
    linkedin = kirtan-kunwar
]
\fnmark[1]
\ead{kunwarkirtan36@gmail.com}
\credit{Methodology (network design), Validation, Writing – original draft, 
        Writing – review \& editing, Results analysis}

\author[1]{Navin Nepal}[
    orcid    = 0009-0007-8733-3826,
    linkedin = navin
]
\fnmark[1]
\ead{navinkhanal3@gmail.com}
\credit{Training, Formal analysis, Literature review, 
        Writing – original draft (sections), Writing – review \& editing}

\author[1]{Dinesh Baniya Kshatri}[
]
\ead{dinesh@ioe.edu.np}
\credit{Supervision, Project administration, Writing – review}

\affiliation[1]{organization={Department of Electronics and Computer Engineering, Thapathali Campus, Tribhuwan University},
            addressline={}, 
            city={Kathmandu},
            postcode={46000}, 
            country={Nepal}}

\cortext[1]{Corresponding author}

\fntext[1]{These authors contributed equally to this work.}


\begin{abstract}
We investigate the use of wavelet-space feature decomposition in neural super-resolution for rendering pipelines. Building on recent neural upscaling frameworks, we introduce a formulation that predicts stationary wavelet coefficients rather than directly regressing RGB values. This frequency-aware decomposition separates low- and high-frequency components, enabling sharper texture recovery and reducing blur in challenging regions. Unlike conventional wavelet transforms, our use of the stationary wavelet transform (SWT) preserves spatial alignment across subbands, allowing the network to integrate G-buffer attributes and temporally warped history frames in a shift-invariant manner. The predicted coefficients are recombined through inverse wavelet synthesis, producing resolution-consistent reconstructions across arbitrary scale factors. We conduct extensive evaluations and ablations, showing that incorporating SWT improves both fidelity and perceptual quality with only modest overhead, while remaining compatible with standard rendering architectures. Taken together, our results suggest that wavelet-domain neural super-resolution provides a principled and efficient path toward higher-quality real-time rendering, with broader implications for neural rendering and graphics applications.
\end{abstract}





\maketitle

\section{Introduction}\label{sec:Introduction}

The demand for high-resolution graphics at real-time frame rates continues to push modern rendering pipelines to their computational limits. Physically based shading and ray tracing greatly improve realism but are often too expensive to run at native resolutions. Image-space super-resolution has therefore become an essential component of real-time graphics systems: frames are rendered at lower resolution and then reconstructed to display resolution, trading computation for learned upscaling.

Neural approaches have recently shown strong potential in this space. Methods such as NSRR~\cite{NSRR:2020}, FuseSR~\cite{FuseSR:2023}, and DFASR~\cite{DFASR:2024} combine low-resolution shading with auxiliary scene information such as G-buffers and motion vectors to reconstruct temporally stable images. DFASR in particular demonstrated arbitrary-scale super-resolution by leveraging Fourier-based implicit neural representations (INRs). Building on this line of work, we ask whether alternative spectral representations could further enhance reconstruction quality.

We propose a wavelet-domain formulation for neural super-resolution tailored to real-time rendering pipelines. Rather than directly predicting pixel intensities, our model regresses wavelet coefficients that encode localized frequency information. This explicit spectral representation improves recovery of high-frequency details and reduces blur in textured regions. To avoid resolution loss across subbands, we employ the stationary wavelet transform (SWT), with inverse synthesis embedded into the network for end-to-end training. Our architecture retains the arbitrary-scale capability of DFASR but achieves sharper detail and improved temporal stability by operating in coefficient space. While absolute performance depends on hardware, our formulation provides a practical path toward higher-fidelity reconstruction within the constraints of real-time rendering pipelines.

\section{Related Work}\label{sec:RelatedWork}

This section surveys prior work most relevant to our setting. We begin with neural super-resolution for real-time rendering, focusing on methods that leverage engine metadata, high-resolution G-buffers, and coordinate-based predictors. We then review wavelet-based super-resolution, contrasting discrete and stationary transforms and common strategies that supervise in coefficient and image space. Finally, we summarize BRDF demodulation and evaluation protocols, including perceptual metrics, which frame our experimental design and analysis.

\subsection{Super-Resolution for Realtime-Rendering}
Early neural approaches to real-time upsampling leveraged \emph{engine metadata} (depth, motion vectors) with temporally-aware networks to reconstruct high-resolution frames from low-resolution inputs. Neural Supersampling (NSRR) exemplifies this, formulating reconstruction as a temporally conditioned enhancement task and achieving stable \(4\times\) upsampling using dense motion and depth cues \cite{NSRR:2020}. While temporally coherent, these methods remain limited by the LR input, struggling to recover high-frequency detail and typically operate at fixed scale factors \cite{NSRR:2020}.

Later work incorporates \emph{high-resolution G-buffers} to provide geometric and material detail efficiently. FuseSR illustrates this trend, using a multi-resolution fusion architecture (“H-Net”) to align and fuse LR color with HR G-buffers, enabling temporally consistent \(4\times\) and \(8\times\) upsampling at 4K \cite{FuseSR:2023}. This mitigates LR information limits but still depends on discrete scales and accurate multi-resolution alignment.

DFASR advances arbitrary-scale reconstruction with a Fourier-mapped implicit representation, predicting pixel values at requested coordinates while using auxiliary modules for detail and robustness \cite{DFASR:2024}. This removes fixed-scale constraints but introduces sensitivity to Fourier parameterization and frequency–band trade-offs \cite{DFASR:2024}. Complementary methods like ExtraNet focus on \emph{temporal} supersampling, extrapolating future frames with HR G-buffers and motion cues to reduce latency without look-ahead \cite{ExtraNet:2021}.

\subsection{Wavelets for Super-Resolution}
Wavelet-based super-resolution operates in a multi-band frequency domain, separating coarse structure from directional details. Early CNNs like Deep Wavelet Prediction regress subband coefficients and invert the transform, improving high-frequency recovery over direct RGB regression \cite{DWSR:2017}. Later variants integrate convolutional backbones with wavelet analysis/synthesis blocks, achieving consistent gains across SR benchmarks \cite{WaveletTransformSR:2021}.

Two common transforms are the discrete wavelet transform (DWT) and the stationary wavelet transform (SWT). DWT is computationally lighter but shift-variant due to downsampling, which can make learning sensitive to pixel shifts. SWT, in contrast, removes downsampling, producing shift-invariant, full-resolution subbands that preserve fine details across spatial locations, at the cost of higher memory and compute \cite{SWT:2022}. Most works use single-level decomposition and supervise both coefficient and image space to stabilize training and retain high-frequency fidelity \cite{DWSR:2017,WaveletTransformSR:2021}. Common orthogonal families (Haar/db1, Daubechies) are often employed, and the transform choice is typically treated as a modular trade-off between reconstruction quality and runtime efficiency.

\subsection{BRDFs and Evaluation Metrics}
Radiance demodulation separates smooth lighting from high-resolution material detail, applying SR on lighting and re-modulating with HR materials for sharp textures and temporal stability \cite{NSSRRD:2024}. Relatedly, pre-integrated BRDFs stabilize filtering/reconstruction \cite{BRDF:2021}, and FuseSR shows combining LR color with HR G-buffers plus BRDF demodulation improves \(4\times\)–\(8\times\) upsampling \cite{FuseSR:2023}. DFASR targets arbitrary-scale reconstruction via Fourier-based implicit representation and spatial–temporal masks rather than BRDF factorization \cite{DFASR:2024}.

Evaluation commonly uses PSNR/SSIM, with LPIPS increasingly adopted to better reflect perceptual quality \cite{LPIPS:2018}. HDR metrics exist but are less standardized in real-time SR, so many works report tone-mapped results for comparability.

\section{Method}
\label{sec:methodoloty}

Our framework extends recent rendering-aware neural super-resolution models by shifting the prediction domain from RGB pixels to wavelet coefficients. While our architecture follows the general coordinate-based design of DFASR~\cite{DFASR:2024}, our key distinction lies in formulating super-resolution as a wavelet reconstruction task. By predicting frequency-aware subbands and reconstructing through the stationary wavelet transform (SWT), we preserve spatial resolution across coefficients and improve recovery of fine detail. The following subsections introduce the problem setup, data preprocessing pipeline, and network architecture, culminating in our integration of wavelet-domain reconstruction.

\subsection{Overview}
\label{sec:method-overview}

Our method formulates real-time super-resolution as a wavelet-domain reconstruction problem. The goal is to recover a high-resolution frame $\hat{I}_{HR}$ from a combination of low-resolution shading, multi-resolution G-buffers, and temporal history. We denote the overall process as
\[
\hat{I}_{t}^{HR} = \mathrm{SR}(I_{t}^{LR}, G_{t}^{LR}, G_{t}^{HR}, I_{t-1}^{HR}, G_{t-1}^{HR}, MV_{t}, F),
\]
where $I_{t}^{LR}$ is the current low-resolution frame, $G_{t}^{LR}$ and $G_{t}^{HR}$ are the low- and high-resolution G-buffers, $I_{t-1}^{HR}$ is the previous high-resolution frame, $G_{t-1}^{HR}$ its corresponding G-buffers, $MV_{t}$ the motion vectors, and $F$ the material reflectance information. This compact formulation captures the pipeline-level data flow, while the details of each component are discussed in later subsections.

Unlike prior approaches that directly regress RGB values, our network predicts a wavelet representation of the target frame. Specifically, it estimates subband coefficients $\hat{W} = \{\hat{C}_{LL}, \hat{C}_{LH}, \hat{C}_{HL}, \hat{C}_{HH}\}$, which are subsequently combined through an inverse stationary wavelet transform:
\[
\hat{I}_{t}^{HR} = \mathcal{W}^{-1}(\hat{W}).
\]
This shift to the frequency domain is particularly well-suited for rendering data: low-frequency lighting and high-frequency material details are naturally separated across subbands, making it easier for the network to reconstruct fine edges, specular highlights, and shading-dependent detail. The result is an upscaling pipeline that aligns more closely with the underlying rendering process while improving fidelity at larger scaling factors.

\subsection{Wavelet-Space Formulation}
\label{sec:wavelet_formulation}

We cast super resolution as prediction in the wavelet domain rather than RGB space. 
Let $\mathcal{W}_{\tau,b}$ denote a single-level 2D wavelet analysis operator with transform type $\tau \in \{\text{DWT}, \text{SWT}\}$ and orthogonal basis $b$ (for example, Haar, DB4, Sym4). 
For a target high-resolution image $I_{t}^{HR} \in \mathbb{R}^{(sH)\times(sW)\times 3}$, the corresponding subbands are
\[
W_{t} \triangleq \big\{ C_{LL}, C_{LH}, C_{HL}, C_{HH} \big\} 
\;=\; \mathcal{W}_{\tau,b}\!\left(I_{t}^{HR}\right),
\]
where $s$ is the requested scale factor. 
The network predicts $\widehat{W}_{t}=\{\widehat{C}_{LL}, \widehat{C}_{LH}, \widehat{C}_{HL}, \widehat{C}_{HH}\}$ directly, and the output image is obtained by inverse synthesis,
\[
\widehat{I}_{t}^{HR} \;=\; \mathcal{W}^{-1}_{\tau,b}\!\left(\widehat{W}_{t}\right).
\]

We use a single level of decomposition, which balances frequency separation against runtime. 
Each color channel is decomposed into four subbands, hence twelve coefficient maps in total. 
For SWT, subbands preserve spatial size, that is $C_{\ast} \in \mathbb{R}^{(sH)\times(sW)\times 3}$, which simplifies alignment across branches and avoids resolution loss inside the network. 
For DWT, subbands are decimated, so we store them in a space-to-depth layout that is convolution friendly and later recover spatial layout with a pixel shuffle prior to inverse synthesis. 
The inverse wavelet transform is implemented inside the computational graph so gradients flow through $\mathcal{W}^{-1}_{\tau,b}$ during training.

We primarily use SWT for its shift invariance and resolution preservation, at a modest compute cost due to non-decimated filtering. 
SWT also requires careful boundary handling. 
To suppress minor edge artifacts from the learned filters and padded convolutions, we crop a one-pixel border when computing losses during training for SWT configurations. 
The basis $b$ is a runtime choice, for example Haar, DB4, or Sym4, and Section~\ref{sec:ablation} reports the effect of $\tau$ and $b$ on accuracy.

\begin{figure}
    \centering
    \includegraphics[width=\linewidth]{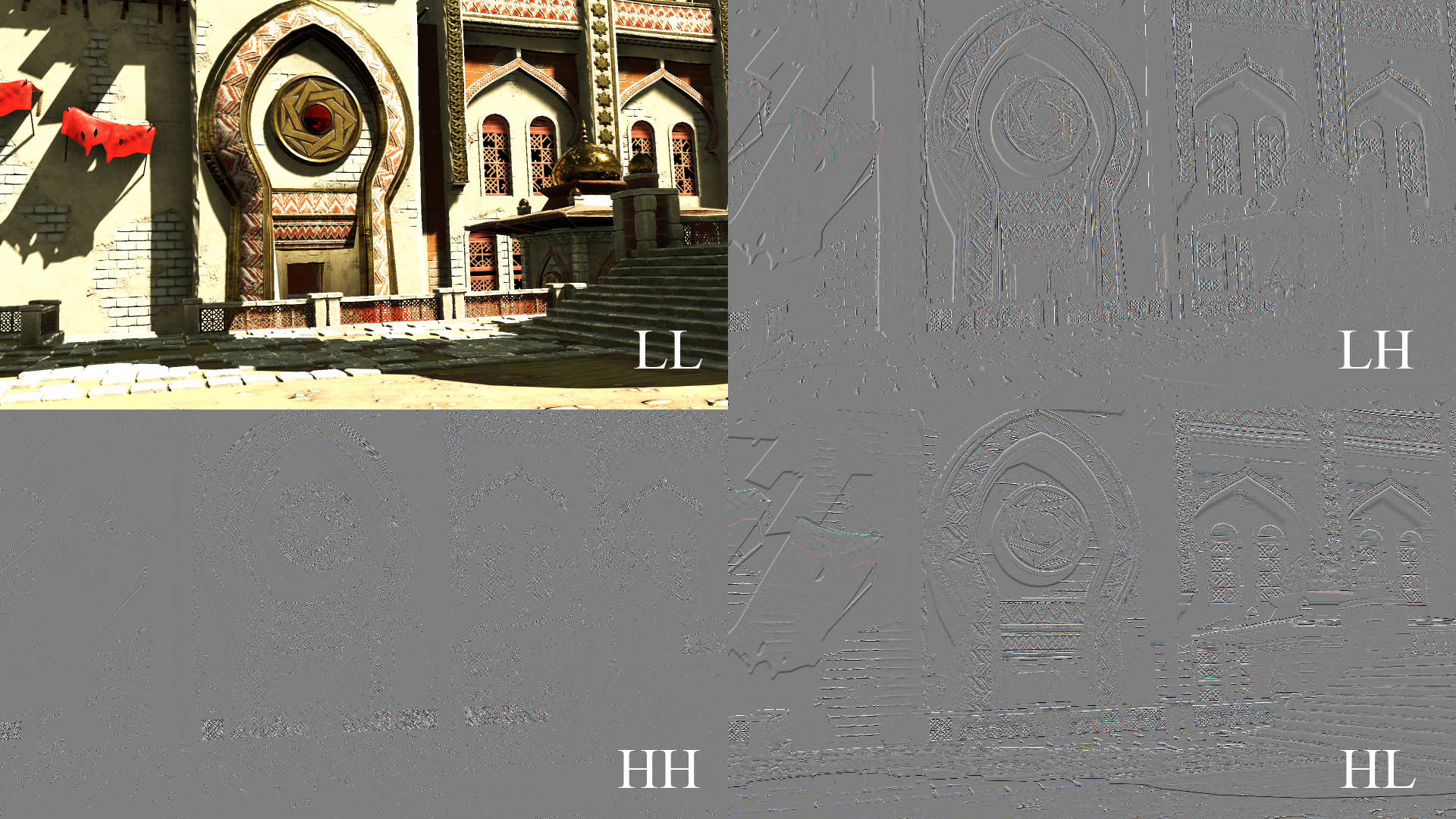}
    \caption{Single-level wavelet decomposition of an RGB image into $LL$, $LH$, $HH$, and $HL$ subbands per channel. 
    Our network predicts these subbands directly, then applies $\mathcal{W}^{-1}_{\tau,b}$ to reconstruct $\widehat{I}_{t}^{HR}$.}
    \label{fig:wavelet_decomposition}
\end{figure}

\subsection{Data Preprocessing and Input Construction}
\label{sec:data-preprocessing}

To align super-resolution with the rendering process, our model operates in \emph{irradiance space}, where shaded images are factorized into BRDF and irradiance components. This decomposition allows the network to focus on reconstructing lighting variation separately from view- and material-dependent reflectance. A low-resolution frame $I_{t}^{LR}$ is demodulated using its per-pixel BRDF term $F_{t}^{LR}$:
\[
L_{t}^{LR} = \frac{I_{t}^{LR}}{F_{t}^{LR}},
\]
where $L_{t}^{LR}$ denotes the demodulated irradiance. The network then predicts the high-resolution irradiance $\hat{L}_{t}^{HR}$:
\[
\hat{L}_{t}^{HR} = \mathcal{M}\!\left(L_{t}^{LR},\, G_{t}^{HR},\, Warp(L_{t-1}^{HR}),\, \Phi\right),
\]
with $\mathcal{M}$ denoting our wavelet-based reconstruction model. Here $G_t^{HR}$ is the high-resolution G-buffers, $Warp(\cdot)$ is temporal reprojection of the previous irradiance frame using motion vectors $MV_t$, and $\Phi = \{\Phi_{\text{spat}}, \Phi_{\text{temp}}\}$ are spatial and temporal masks as introduced in \cite{DFASR:2024}.

The G-buffers contain albedo ($A$), normals ($N$), depth ($D$), roughness ($R$), and metallic ($M$), from which per-pixel BRDF factors ($F = \{F_t^{LR}, F_t^{HR}, F_{t-1}^{HR}\}$) are derived. After super-resolution, the final shaded output is obtained by remodulation:
\[
\hat{I}_{t}^{HR} = F_{t}^{HR} \cdot \hat{L}_{t}^{HR},
\]
which restores view- and lighting-dependent effects. This formulation shifts the network’s focus toward reconstructing frequency content in irradiance space, while high-resolution geometric cues from $G_t$ ensure accurate remodulation in the final image.

\begin{figure}
\centering
\begin{minipage}[b]{0.32\linewidth}
    \centering
    \includegraphics[width=\linewidth]{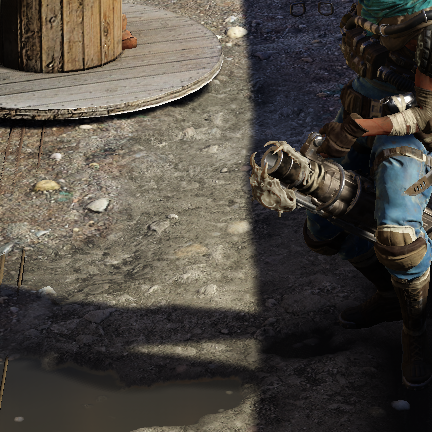}
    \vspace{2pt}
    \small Rendered Frame
\end{minipage}
\begin{minipage}[b]{0.32\linewidth}
    \centering
    \includegraphics[width=\linewidth]{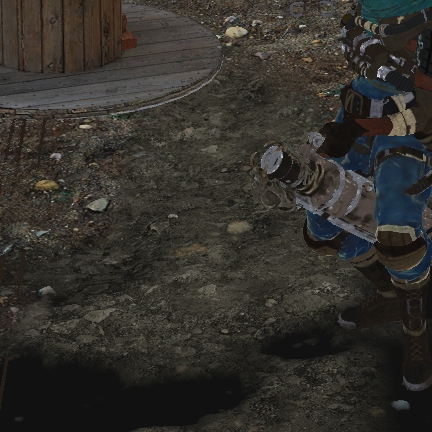}
    \vspace{2pt}
    \small BRDF
\end{minipage}
\begin{minipage}[b]{0.32\linewidth}
    \centering
    \includegraphics[width=\linewidth]{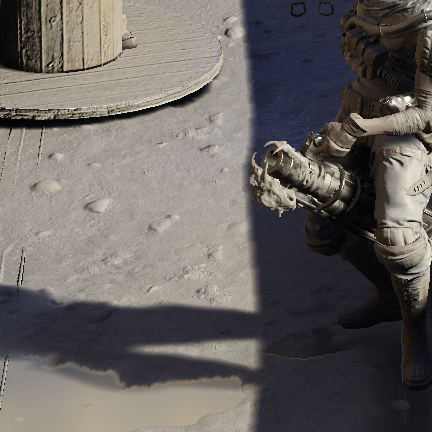}
    \vspace{2pt}
    \small Irradiance
\end{minipage}
\caption{Illustration of BRDF demodulation. The shaded frame is factorized into BRDF and irradiance. This separation shifts the network’s learning focus to low-frequency lighting while high-frequency reflectance patterns are reapplied during remodulation.}
\label{fig:brdf_demo}
\end{figure}

\subsection{Network Architecture}
\label{sec:network-architecture}

\begin{figure*}
    \centering
    \includegraphics[width=0.95\linewidth]{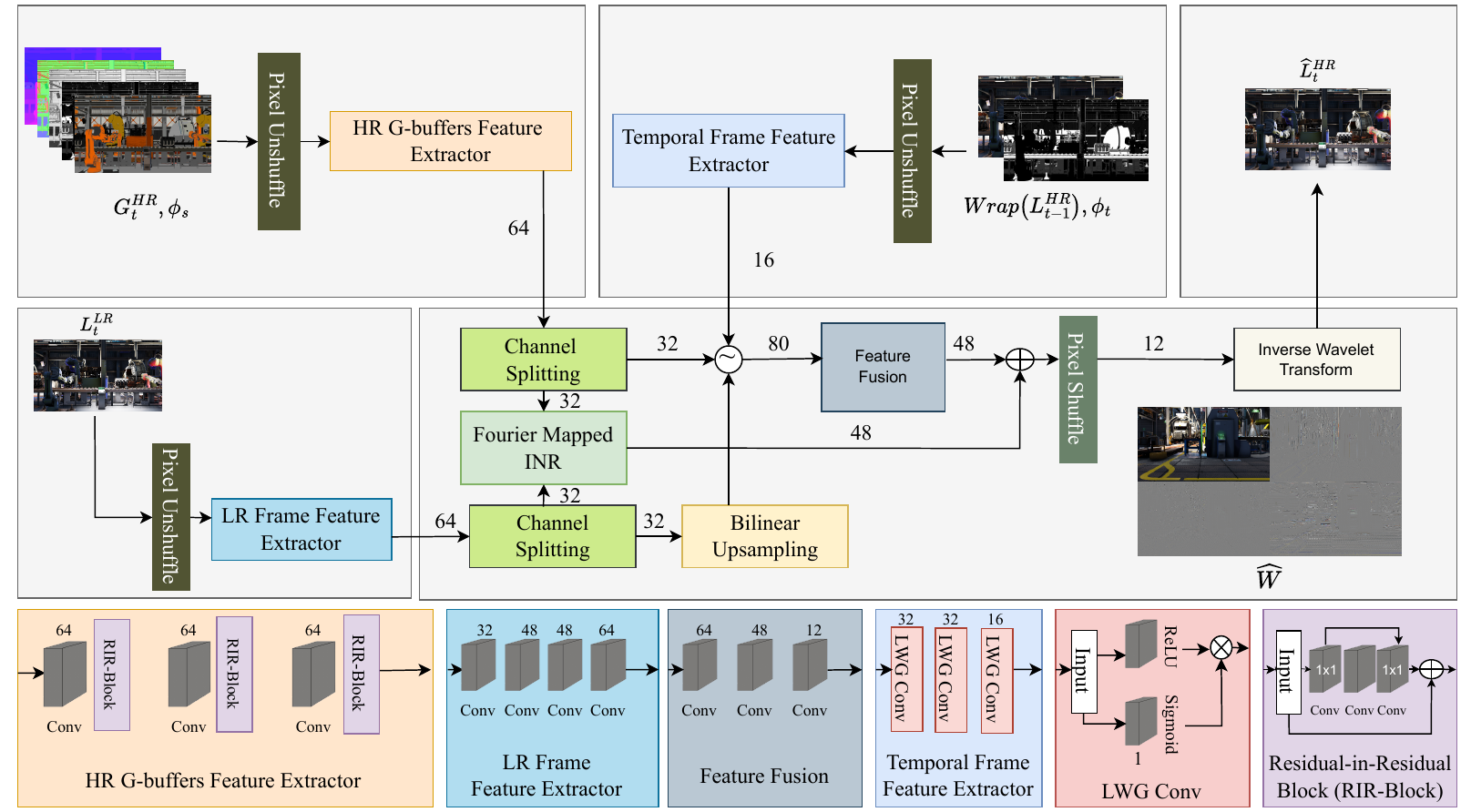}
    \caption{Overview of our network architecture. The model consists of three feature extractors (for LR frames, HR G-buffers, and temporal inputs), a convolutional fusion branch, and an implicit neural representation (INR) branch with Fourier feature mapping. Their outputs are combined in the wavelet domain and reconstructed into the final image via inverse wavelet transform and pixel shuffle. Convolutional blocks are $3\times3$ with padding 1, except the RIR-blocks which consist of $1\times1$ expand/reduce layers with an inner $3\times3$ convolution.}
    \label{fig:arch}
\end{figure*}

Our model adapts the DFASR backbone but shifts the output space to wavelet coefficients. As shown in Fig.~\ref{fig:arch}, the network processes inputs through dedicated feature extractors, and combines them via a feature fusion path and a Fourier-mapped INR branch. The combined output is reconstructed into the final image using the inverse wavelet transform.

\subsubsection{Feature Extractors}
\label{sec:feat-extractors}

Our model employs three dedicated feature extractors tailored to the input modalities: 
a low-resolution frame extractor (LR-FE), a high-resolution G-buffer extractor (GB-FE), 
and a temporal extractor (T-FE). Each produces a compact representation aligned to the target resolution 
and contributes complementary cues for wavelet prediction.

The \textbf{LR-FE} processes the pixel-unshuffled, demodulated low-resolution input frame $I_t^{LR}$, 
using a stack of $3 \times 3$ convolutions to produce a feature map later routed to both fusion and INR pathways.  
The \textbf{GB-FE} encodes the high-resolution G-buffers $G_t^{HR}$ through RIR blocks, extracting geometry- and 
material-aware information such as depth and surface orientation at display resolution.  
The \textbf{T-FE} receives the warped previous frame $\mathrm{Warp}(I_{t-1}^{HR}, MV_t)$ together with temporal masks, 
and is implemented with lightweight gated convolutions (LWGC) to balance motion sensitivity and efficiency.

Together, these extractors provide complementary signals: LR features represent coarse appearance, 
HR G-buffers supply high-frequency geometric guidance, and temporal features capture frame-to-frame coherence. 
Their outputs serve as the basis for subsequent fusion and wavelet-domain reconstruction.

\subsubsection{Feature Fusion}
\label{sec:feature-fusion}

The feature fusion module aggregates information from three complementary sources: low-resolution frame features, high-resolution G-buffer features, and temporally aligned features from the previous frame. This combined representation captures both spatial and geometric context, providing the structural backbone of the reconstruction. The fusion branch is responsible for generating the coarse wavelet-domain prediction, while the implicit neural representation (INR) branch contributes finer local corrections. Their outputs are combined element-wise to form the final set of predicted wavelet coefficients, which are later synthesized into the high-resolution image.

\subsubsection{Implicit Neural Representation (INR)}
\label{sec:inr}

To support arbitrary-scale super-resolution while enhancing localized high-frequency detail, we employ a coordinate-based Implicit Neural Representation (INR) branch. Prior works such as DFASR~\cite{DFASR:2024} demonstrated the effectiveness of Fourier-mapped INRs for rendering-aware upscaling. Building on this idea, we adapt the design to predict wavelet-domain features rather than RGB pixels, which provides finer frequency separation and improved detail recovery.

For each HR pixel coordinate $x$, we compute a Fourier embedding
\[
\mathcal{F}(x) = A(x) \cdot
\begin{bmatrix}
\cos(\pi \langle F(x), x \rangle + \phi(x)) \\
\sin(\pi \langle F(x), x \rangle + \phi(x))
\end{bmatrix},
\]
where $A(x)$ is an amplitude term derived from LR features, $F(x)$ is a frequency term guided by HR G-buffer features, and $\phi(x)$ is a spatial phase offset. These embeddings are passed through a lightweight multi-layer perceptron (MLP), producing predictions in wavelet space.

The INR complements the fusion module by injecting coordinate-aware, high-frequency corrections. Their outputs are combined element-wise to form the final set of predicted coefficients. Together, the two branches balance structural fidelity with detail sharpness, while the continuous INR mapping enables reconstructions at arbitrary scale factors.

\subsubsection{Reconstruction}
\label{subsec:reconstruction}

The final reconstruction stage aggregates the outputs from the feature fusion and INR branches, which are element-wise combined to form the estimated wavelet coefficients. These coefficients are subsequently passed through a reconstruction layer consisting of a lightweight convolution and a single-level inverse wavelet transform (IWT), following the formulation already introduced in Sec.~\ref{sec:wavelet_formulation}. This yields the super-resolved output image.

Our implementation supports configurable reconstruction settings: the wavelet basis can be selected at runtime (e.g., Haar, Daubechies), and both discrete (DWT) and stationary (SWT) variants are available. By default, the reconstruction is performed with a single-level discrete transform, but the flexibility of our framework allows controlled experimentation with different bases and transform types.

The resulting output is produced in HDR linear space. For visualization purposes, we apply tonemapping during evaluation, but training and inference operate directly in the HDR domain to preserve photometric consistency.

\subsubsection{Loss Functions}
\label{sec:loss}

Training is guided by a composite objective that balances fidelity in both the image and wavelet domains. We combine (i) an $\ell_1$ loss on predicted wavelet coefficients, (ii) an $\ell_1$ reconstruction loss in image space, (iii) perceptual similarity terms using SSIM and LPIPS, and (iv) spatial mask and temporal consistency losses following DFASR~\cite{DFASR:2024}. The total loss is
\[
\begin{aligned}
\mathcal{L}_{total} = \;&
\lambda_w \mathcal{L}_{waveletL1} +
\lambda_i \mathcal{L}_{imageL1} +
\lambda_s \mathcal{L}_{SSIM} \\
&+
\lambda_p \mathcal{L}_{LPIPS} +
\lambda_m \mathcal{L}_{mask} +
\lambda_t \mathcal{L}_{temporal}
\end{aligned}
\]
This formulation enforces accuracy across spectral, perceptual, and temporal dimensions, ensuring stable reconstructions under real-time rendering.

\section{Results}
\label{sec:results}

\subsection{Evaluation Setup}
\label{sec:evaluation-setup}

We evaluate our method on a custom Unreal Engine 4.27 dataset containing diverse indoor and outdoor scenes, including urban environments, industrial interiors, and stylized game assets. The dataset comprises approximately 3700 training frames, 700 validation frames, and 1000 test frames, drawn from seven distinct scenes, details in Table~\ref{tab:dataset}. High-resolution (HR) frames are rendered at 1920$\times$1080 resolution along with their corresponding high-resolution G-buffers. To generate low-resolution (LR) inputs, we apply nearest-neighbor downsampling at random scales between $2.0\times$ and $4.0\times$ during training, enabling arbitrary-scale inference. Dataset extraction was facilitated by a custom Unreal Engine plugin, adapted from the Extranets engine modification \cite{ExtraNet:2021}, allowing automated capture of both shading outputs and auxiliary buffers.

\begin{table}
\centering
\caption{Dataset composition with train/validation/test split for each scene. All frames are rendered at 1920$\times$1080 with HR G-buffers and downsampled for LR inputs.}
\label{tab:dataset}
\begin{tabular}{lccc}
\hline
\textbf{Scene} & \textbf{Train} & \textbf{Val} & \textbf{Test} \\
\hline
Eastern Village & 555 & 100 & 285 \\
Abandoned Factory & 740 & 100 & 0 \\
Brass City & 740 & 100 & 0 \\
City Park & 740 & 100 & 0 \\
Factory Interior & 555 & 100 & 185 \\
Slay & 370 & 100 & 185 \\
Downtown Alley & 0 & 100 & 185 \\
Downtown West & 0 & 0 & 285 \\
\hline
\textbf{Total} & 3700 & 700 & 1000 \\
\hline
\end{tabular}
\end{table}

\subsection{Quantitative Comparison}
\label{sec:quant-comparison}

We compare our proposed wavelet-space super-resolution framework against three representative baselines: (i) classical interpolation (Bicubic), (ii) general-purpose image SR (Real-ESRGAN \cite{RealESRGAN:2021}), and (iii) rendering-oriented neural SR methods (DFASR). Evaluation is conducted at $3\times$ upscaling, reporting PSNR, SSIM, and LPIPS, as summarized in Table~\ref{tab:methods_comp}.

Bicubic serves as a simple baseline but produces heavily smoothed images that lack high-frequency recovery. Real-ESRGAN provides sharper results than Bicubic, but its natural-image prior fails to generalize to rendered frames, often hallucinating textures inconsistent with scene geometry. In contrast, rendering-aware approaches (DFASR and our method) leverage G-buffers and BRDF demodulation to achieve sharper and more stable reconstructions.

Table~\ref{tab:quantitative} reports average performance of different wavelet configurations across test scenes. Our best-performing variant, \textbf{SWT with DB4 wavelet}, achieves the highest PSNR and SSIM and the lowest LPIPS across scales. Compared to the ``No Wavelet'' baseline, it improves PSNR by \textbf{1.5 dB} and reduces LPIPS by \textbf{0.022} on average. While SWT DB4 incurs slightly higher runtime cost than DWT variants (see Section~\ref{sec:runtime}), it offers the best overall trade-off between accuracy and efficiency.

We report Peak Signal-to-Noise Ratio (PSNR), Structural Similarity Index (SSIM), and Learned Perceptual Image Patch Similarity (LPIPS) \cite{LPIPS:2018}. PSNR and SSIM are computed in normalized HDR space with gamma correction and clipping, while LPIPS is evaluated with AlexNet during training and VGG for final reporting. Metrics are averaged across scales of $2\times$, $3\times$, and $4\times$ for each test scene.

All models are trained end-to-end with a patch size of $432 \times 432$ and batch size of 8. Evaluations are performed on full-resolution frames to ensure fidelity across the entire image.

\begin{table*}
\centering
\caption{Comparison with Bicubic, Real-ESRGAN, and DFASR on test scenes (3$\times$ upscaling). Best results in bold. Our SWT DB4 configuration consistently achieves the best performance.}
\label{tab:methods_comp}
\begin{tabular}{lccccccccc}
\toprule
\multirow{2}{*}{Method} &
\multicolumn{3}{c}{Eastern Village} &
\multicolumn{3}{c}{Downtown Alley} &
\multicolumn{3}{c}{Downtown West} \\
\cmidrule(lr){2-4} \cmidrule(lr){5-7} \cmidrule(lr){8-10}
 & PSNR$\uparrow$ & LPIPS$\downarrow$ & SSIM$\uparrow$ & PSNR$\uparrow$ & LPIPS$\downarrow$ & SSIM$\uparrow$ & PSNR$\uparrow$ & LPIPS$\downarrow$ & SSIM$\uparrow$ \\
\midrule
Bicubic & 23.9 & 0.352 & 0.760 & 30.5 & 0.326 & 0.837 & 23.5 & 0.420 & 0.678 \\
Real-ESRGAN & 24.7 & 0.305 & 0.760 & 30.1 & 0.251 & 0.825 & 24.4 & 0.326 & 0.678 \\
\midrule
DFASR & 28.2 & 0.180 & 0.850 & 32.5 & 0.152 & 0.903 & 26.7 & 0.250 & 0.763 \\
SWT DB4 (Ours) & \textbf{29.7} & \textbf{0.115} & \textbf{0.888} & \textbf{36.8} & \textbf{0.059} & \textbf{0.960} & \textbf{28.5} & \textbf{0.144} & \textbf{0.872} \\
\bottomrule
\end{tabular}
\end{table*}

\begin{table}
\centering
\caption{Quantitative comparison of methods (3x upscaling). Best results in bold.}
\label{tab:quantitative}
\begin{tabular}{lccc}
\toprule
Method & PSNR$\uparrow$ & LPIPS$\downarrow$ & SSIM$\uparrow$ \\
\midrule
No Wavelet & 28.9 & 0.136 & 0.897 \\
DWT Haar & 29.4 & 0.124 & 0.905 \\
SWT Haar & 29.7 & 0.117 & 0.912 \\
SWT DB4 & \textbf{30.4} & \textbf{0.114} & \textbf{0.920} \\
\bottomrule
\end{tabular}
\end{table}

\subsection{Qualitative Comparison}
\label{sec:qualitative}

While quantitative metrics such as PSNR, SSIM, and LPIPS provide useful averages, they do not always capture perceptual differences. We therefore present visual comparisons against key baselines, focusing on regions with fine textures, sharp edges, and high-frequency structures.

\begin{figure*}
    \centering
    \includegraphics[width=.98\linewidth]{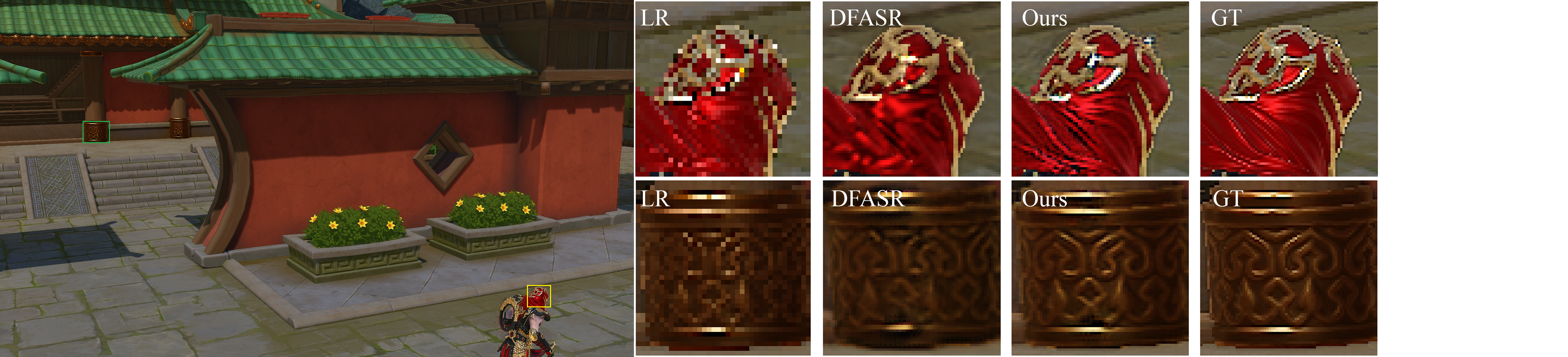}\\[2mm]
    \includegraphics[width=.98\linewidth]{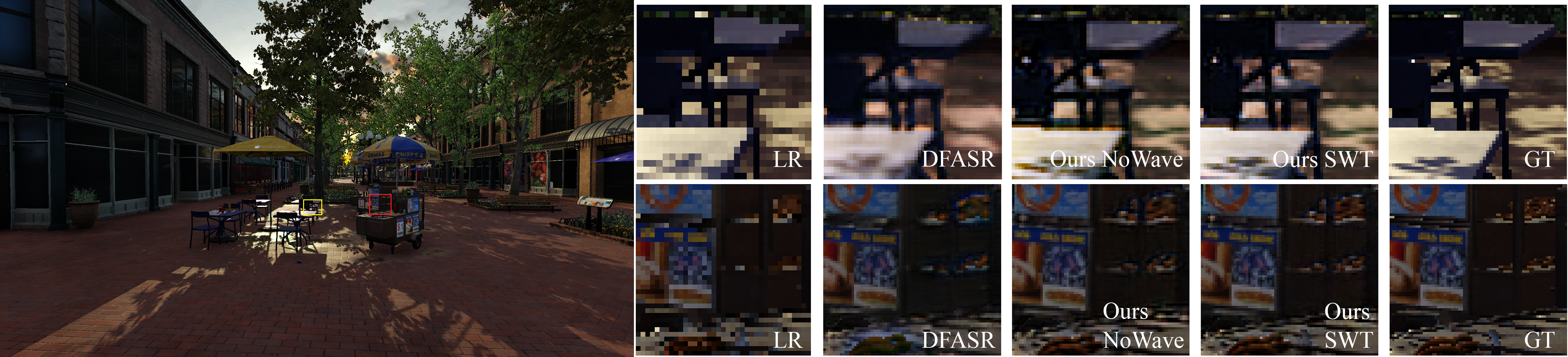}\\[2mm]
    \includegraphics[width=.98\linewidth]{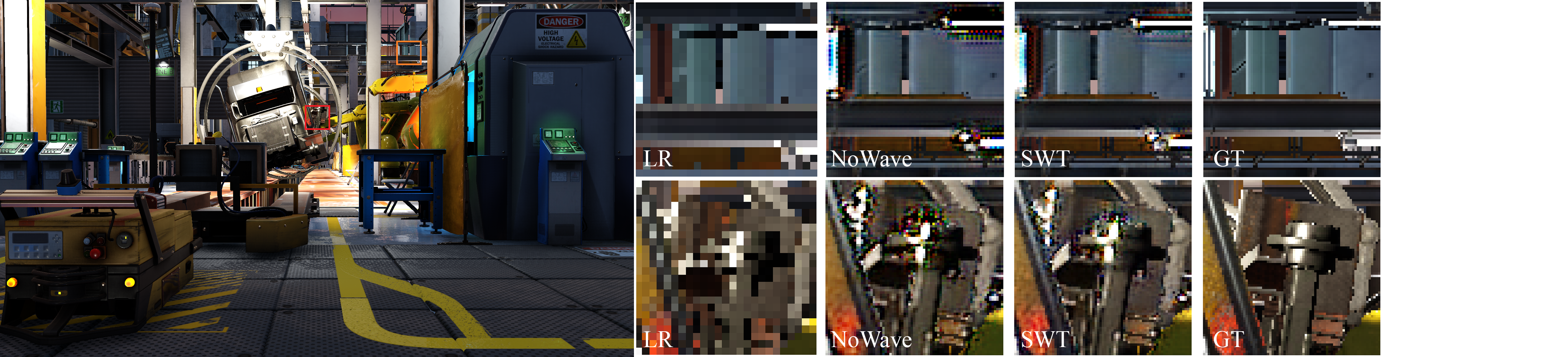}\\[2mm]
\caption{Upscaling results on multiple scenes using NoWave, DFASR \cite{DFASR:2024}, and our SWT model. The first scene (Asian Village) demonstrates that our SWT model recovers sharper details on the head region, where specular highlights are preserved more faithfully compared to DFASR. In the second scene (Downtown West), our method produces clearer structures on the pillars and achieves more accurate color reconstruction, while also maintaining sharper edges in fine elements such as tables. The third scene (Factory) highlights the advantage of SWT in handling challenging color regions, where our model more reliably reconstructs white surfaces that NoWave tends to degrade. Overall, SWT consistently preserves edges and fine details better than the baselines, producing sharper, cleaner, and more realistic outputs across both $2\times$ (first two scenes) and $3\times$ (third scene) upscaling.}
\label{fig:qual_comp_vertical}
\end{figure*}

Classical interpolation (Bicubic) produces overly smoothed reconstructions with little recovery of fine detail, while Real-ESRGAN \cite{RealESRGAN:2021} generates sharper results but still leaves high-frequency textures blurry and occasionally hallucinates patterns inconsistent with geometry. Rendering-aware methods show clearer advantages: the no-wavelet baseline (NoWave) and DFASR preserves overall structure and decent sharpness but often introduces ringing or residual blur. 
Our SWT-based model provides modest but consistent improvements in edge clarity and texture fidelity, 
particularly in repeated patterns, foliage, and diagonal structures. As shown in Figure~\ref{fig:qual_comp_vertical}, SWT restores edges more faithfully than both DFASR and NoWave, 
and in several scenes demonstrates improved color fidelity and reduced errors, resulting in more visually consistent reconstruction.

These visual observations align with the quantitative trends, indicating that the wavelet-space formulation provides a small but reliable improvement in high-frequency detail reconstruction.

\subsection{Ablation Study}
\label{sec:ablation}

To better understand the contributions of different design choices, 
we group ablations into two categories: 
\textit{wavelet formulation} and \textit{architectural/training variants}. 
All results are reported at $2\times$ upscaling on tone-mapped outputs.

\paragraph{Wavelet formulation.}
We compare discrete wavelet transform (DWT) against stationary wavelet transform (SWT), 
and evaluate different wavelet bases. 
This isolates the effect of transform type and basis on reconstruction quality.

\begin{figure}
\centering
\begin{tikzpicture}
\begin{axis}[
    ybar,
    bar width=0.6cm,
    width=0.9\linewidth,
    height=6cm,
    ylabel={PSNR (dB)},
    symbolic x coords={
        No wavelet,
        DWT--Haar,
        SWT--Haar,
        SWT--Sym4,
        SWT--DB4
    },
    xtick=data,
    x tick label style={rotate=30,anchor=east},
    nodes near coords,
    nodes near coords align={vertical},
    every node near coord/.append style={yshift=2pt}, 
    ymin=28, ymax=31,
    axis line style={draw=none}, 
    tick style={draw=none}, 
    major grid style={dashed,gray!30},
    ymajorgrids
]
\addplot[fill=gray!60] coordinates {
    (No wavelet,29.0) 
    (DWT--Haar,29.5) 
    (SWT--Haar,29.9) 
    (SWT--Sym4,30.4) 
    (SWT--DB4,30.5)
};
\end{axis}
\end{tikzpicture}
\caption{PSNR comparison of ablation configurations at $2\times$ upscaling. Values correspond to Table~\ref{tab:ablation_wavelet}. We plot PSNR only for clarity, as SSIM and LPIPS follow the same trend.}
\label{fig:ablation_bar}
\end{figure}
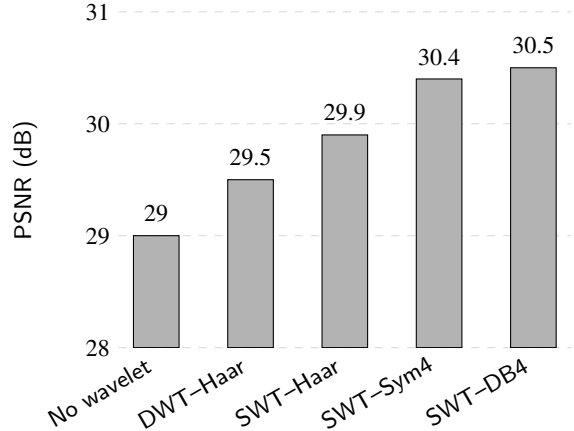

\begin{table}
\centering
\caption{Wavelet formulation ablation at $2\times$ upscaling. 
Metrics are computed on tone-mapped outputs. Best results are highlighted in bold.}
\label{tab:ablation_wavelet}
\begin{tabular}{lccc}
\toprule
Configuration & PSNR $\uparrow$ & LPIPS $\downarrow$ & SSIM $\uparrow$ \\
\midrule
No wavelet (baseline) & 29.0 & 0.135 & 0.899  \\
DWT–Haar & 29.5 & 0.122 & 0.906  \\
SWT–Haar & 29.9 & 0.114 & 0.913  \\
SWT–Sym4 & 30.4 & 0.116 & 0.921  \\
SWT–DB4 & \textbf{30.5} & \textbf{0.112} & \textbf{0.922} \\
\bottomrule
\end{tabular}
\end{table}

\paragraph{Architectural and training variants.}
We further investigate alternative network designs:  
(1) \textbf{Fusion$\rightarrow$LL, INR\allowbreak$\rightarrow$\allowbreak others (SWT\allowbreak--\allowbreak Haar)} restricts the fusion path 
to predict the low-frequency LL subband, while the INR predicts the high-frequency subbands (LH, HL, HH).  
(2) \textbf{Multi-INR (DWT\allowbreak--\allowbreak Haar)} assigns a separate INR branch to each subband instead of a shared INR, 
testing whether coefficient-wise specialization improves results.  
(3) \textbf{Single-scale (2$\times$ only, DWT\allowbreak--\allowbreak Haar)} trains the model exclusively on $2\times$ upscaling 
instead of arbitrary-scale sampling, measuring whether multi-scale training compromises single-scale fidelity.

\begin{table}
\centering
\caption{Architectural and training ablations at $2\times$ upscaling. 
Each variant is evaluated with its native wavelet transform.}
\label{tab:ablation_arch}
\begin{tabular}{lccc}
\toprule
Configuration & PSNR $\uparrow$ & LPIPS $\downarrow$ & SSIM $\uparrow$ \\
\midrule
SWT-Haar & 29.9 & 0.114 & 0.913 \\
Fusion-LL (SWT-Haar) & 29.7 & 0.120 & 0.910  \\
\midrule
DWT-Haar & 29.5 & 0.122 & 0.906 \\
Multi-INR (DWT–Haar) & 29.4 & 0.126 & 0.905  \\
Single-scale (DWT–Haar) & 29.7 & 0.123 & 0.909  \\
\bottomrule
\end{tabular}
\end{table}

\paragraph{Takeaways.}
Overall, the ablation results show that SWT consistently outperforms DWT, 
with the DB4 basis giving the highest reconstruction quality among all tested configurations. 
Alternative architectural variants such as Fusion-LL and Multi-INR reduce performance, 
indicating that a unified INR with joint coefficient prediction is more effective. 
Finally, multi-scale training improves generalization without sacrificing fixed-scale accuracy, 
demonstrating the robustness of our chosen setup.

\subsection{Runtime and Efficiency}
\label{sec:runtime}

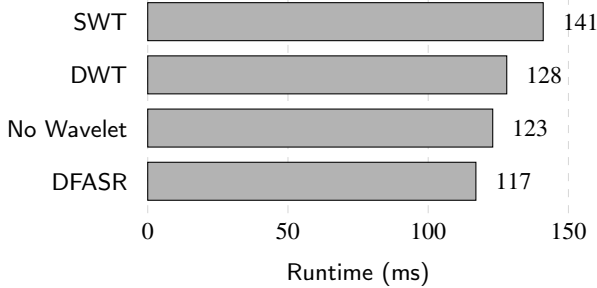
\begin{figure}
\centering
\begin{tikzpicture}
\begin{axis}[
    xbar=0pt, 
    bar width=0.5cm,
    width=0.85\linewidth,
    height=6cm,
    xlabel={Runtime (ms)},
    symbolic y coords={DFASR, No Wavelet, DWT, SWT},
    ytick=data,
    y=0.7cm, 
    enlarge y limits=0.2, 
    nodes near coords,
    nodes near coords align={horizontal},
    every node near coord/.append style={xshift=4pt}, 
    xmin=0, xmax=150,
    axis line style={draw=none}, 
    tick style={draw=none}, 
    ytick style={draw=none}, 
    major grid style={dashed,gray!30},
    xmajorgrids 
]

\addplot [fill=gray!60] coordinates {
    (117,DFASR) 
    (123,No Wavelet) 
    (128,DWT) 
    (141,SWT)
};

\end{axis}
\end{tikzpicture}
\caption{Average per-frame runtime on an RTX 3050 mobile GPU at $1920\times1080$ output ($2\times$ upscaling). Absolute numbers are hardware dependent, but relative overhead remains modest: DWT adds $\sim$5 ms and SWT adds $\sim$18 ms compared to the No-Wavelet baseline.}
\label{fig:runtime_bar}
\end{figure}

We evaluate inference performance at $2\times$ upscaling for $1920\times1080$ output on an RTX 3050 mobile GPU, using single-frame inference without low-level optimizations. Figure~\ref{fig:runtime_bar} reports average per-frame runtimes across methods. Our No-Wavelet baseline is close to DFASR, with a minor discrepancy of $\sim$6 ms, which we attribute to implementation-level differences. Introducing wavelet transforms increases runtime: DWT variants add $\sim$5 ms overhead relative to the No-Wavelet baseline, while SWT adds $\sim$18 ms. This reflects a clear accuracy–efficiency trade-off: SWT with DB4 provides the best perceptual quality but incurs additional runtime cost.

Absolute runtimes on the RTX 3050 mobile are higher ($\sim$141 ms) than DFASR’s published results on the RTX 4090 (11 ms), primarily due to the lack of RT cores and FP16 acceleration on our hardware. Importantly, the relative overhead compared to the DFASR baseline is modest (+24 ms). We therefore expect that on high-end GPUs, our approach would remain real-time capable, consistent with DFASR’s original claims.

\section{Conclusion}
\label{sec: Conclusion}

We have presented a wavelet-domain super-resolution framework for real-time rendering pipelines, predicting frequency-aware subbands rather than directly regressing RGB values. By combining convolutional feature extractors, an implicit neural representation with Fourier mapping, and stationary wavelet reconstruction, our method achieves detail-preserving, resolution-consistent upscaling across arbitrary scales. Experimental results show that this formulation effectively recovers high-frequency detail while remaining compatible with standard rendering pipelines.

Several challenges remain, including limited temporal robustness compared to video SR methods, the use of only a single-level wavelet decomposition, and the need for further optimization to reach practical deployment on mid-range hardware. Future work may address these by exploring multi-level wavelet hierarchies, stronger temporal modeling, and hardware-aware inference strategies. Coupling the framework with real-time ray tracing pipelines also offers a promising direction to exploit richer lighting signals within the same frequency-aware formulation.

\section*{Acknowledgements}

The authors would like to express their sincere gratitude to Er. Saroj Shakya for his invaluable guidance and support as a co-supervisor throughout the duration of this project. The Department of Electronics and Computer Engineering, Thapathali Campus, is gratefully acknowledged for providing the opportunity and resources to carry out this research. The authors also acknowledge Ashim Bhattarai and Ashim Tiwari for their assistance with computational resources.

\section*{Data Availability Statement}
The source code for both the proposed neural network and the Unreal Engine plugin used in the experiments 
is publicly available at \url{https://github.com/Prateek61/WDSS}. 
All results in the paper can be reproduced using the provided code and instructions.


\bibliographystyle{cas-model2-names}
\bibliography{cas-refs}



\end{document}